\documentclass{emulateapj}
\usepackage{apjfonts}

\newcommand{\rvec}{\mbox{\boldmath $r$}}

\newcommand{\zetavec}{\mbox{\boldmath $\zeta$}}

\newcommand{\te}{t_{\rm E}}
\newcommand{\tep}{t_{{\rm E},p}}
\newcommand{\re}{r_{\rm E}}
\newcommand{\retilde}{\tilde{r}_{\rm E}}
\newcommand{\pivec}{\mbox{\boldmath $\pi$}}
\newcommand{\thetae}{\theta_{\rm E}}
\newcommand{\thetaep}{\theta_{{\rm E},p}}
\newcommand{\sn}{{\rm S/N}}

\def\eqalign#1{\null\,\vcenter{\openup\jot
        \ialign{\strut\hfil$\displaystyle{##}$&$
        \displaystyle{{}##}$\hfil \crcr#1\crcr}}\,}

\begin{document}
\title{Microlensing Characterization of Wide-Separation Planets}

\author 
{
Cheongho Han\altaffilmark{1,2},
B.\ Scott Gaudi\altaffilmark{3},
Jin H. An\altaffilmark{4}, and 
Andrew Gould\altaffilmark{2}}
\altaffiltext{1}{Department of Physics, Institute for Basic Science
Research, Chungbuk National University, Chongju 361-763, Korea}
\altaffiltext{2}{Ohio State University, Department of Astronomy, 
Columbus, OH 43210}
\altaffiltext{3}{Harvard-Smithsonian Center for Astrophysics, 
60 Garden St., Cambridge, MA 02138}
\altaffiltext{4}{Institute of Astronomy, University of Cambridge, 
Madingley Road, Cambridge CB3 0HA, UK}
\email{cheongho@astroph.chungbuk.ac.kr, sgaudi@cfa.harvard.edu,
jin@ast.cam.ac.uk, gould@astronomy.ohio-state.edu}


\begin{abstract}
With their excellent photometric precision and dramatic increase in
monitoring frequency, future microlensing survey experiments are
expected to be sensitive to very short time-scale, isolated events
caused by free-floating and wide-separation planets with mass as low
as a few lunar masses.  The scientific value of these detections would
be greatly enhanced if their nature (bound or unbound) could be
accurately characterized, and if the planet masses could be measured.
We estimate the probability of measuring the Einstein radius $\thetae$
for bound and free-floating planets; this is one of the two additional
observables required to measure the planet mass.  We carry out detailed 
simulations of the planetary events expected in next-generation surveys 
and estimate the resulting uncertainty in $\thetae$ for these events.  
We show that, for main-sequence sources and Jupiter-mass planets, the 
caustic structure of wide-separation planets with projected separations 
of $\la 20~{\rm AU}$ substantially increases the probability of measuring 
the dimensionless source size and thus determining $\thetae$ compared 
to the case of unbound planets.  In this limit where the source is much 
smaller than the caustic, the effective cross-section to measure $\thetae$ 
to $10\%$ is $\sim 25\%$ larger than the full width of the caustic.  
Measurement of the lens parallax is possible for low-mass planetary 
events by combined observations from the ground and a satellite located 
in an L2 orbit; this would complete the mass measurements for such 
wide-separation planets.  Finally, short-duration events caused by 
bound planets can be routinely distinguished from those caused by 
free-floating planets for planet-star separations $\la 20~{\rm AU}$ 
from either the deviations due to the planetary caustic or (more often) 
the low-amplitude bump from the magnification due to the parent star.
\end{abstract}


\section{Introduction}
\label{intro}

Microlensing experiments were originally proposed to search for
Galactic dark matter in the form of massive compact objects (MACHOs)
\citep{paczynski86}.  However, microlensing developed several other
applications including the detection and characterization of
extrasolar planets \citep{mao91, gould92}.  Recently, \citet{bond04}
reported the first clearcut microlensing detection of an exoplanet.

Microlensing planet searches currently operate in the survey/follow-up
mode.  Large areas of the sky are sparsely monitored by survey
collaborations \citep{alcock96, udalski01,bond02} to detect ongoing 
microlensing events arising from normal stars.  These events are 
alerted in real-time before the event peak, and then individually 
followed-up with the dense sampling needed to detect the
short-duration (about a day for a Jupiter-mass planet and a few hours
for an Earth-mass planet) perturbation to the primary lightcurve
caused by a planetary companion to the primary lens 
\citep{albrow98, rhie00, yoo04b}.  One limitation of this type of 
planet search strategy, however, is that events are only efficiently 
followed when the source is located within the Einstein ring radius 
of the primary.  Generally, such source positions are only sensitive 
to planets with separations located within a certain range of distances 
from their host stars. Planets in this so called `lensing zone' have 
separations in the range of $0.6\la s \la 1.6$, where $s$ is the 
projected star-planet separation normalized by the Einstein ring radius.  
For a typical Galactic bulge event with a lens and a source located at 
$D_{ol}=6$ kpc and $D_{os}=8$ kpc, respectively, the Einstein ring has 
a radius of
\begin{equation}
\re \sim 2\ {\rm AU}\left( {M\over 0.3\ M_\odot}\right)^{1/2},
\label{eq1}
\end{equation}
where $M$ is the primary-star mass, and thus current microlensing
planet searches are primarily sensitive to planets in the range of
projected physical separations, $r_\perp$, of $1\ {\rm AU}\la r_\perp
\la 5\ {\rm AU}$.  Furthermore, because the follow-up is generally done
with small field-of-view instruments, the events must be monitored
sequentially.  As a result, only a few events can be followed at any
given time, and it is difficult to achieve the requisite temporal
sampling on a sufficient number of events to detect short-duration,
low-probability events such as those caused by low-mass or
large-separation planets.

These limitations can be overcome with the advent of future lensing 
experiments that use very large-format imaging cameras to survey 
wide fields continuously at high cadence.  These next-generation 
surveys dispense with the alert/follow-up mode of searching for 
planets, and instead simultaneously obtain densely-sampled lightcurves 
of all microlensing events in their field-of-view (FOV).  Because all 
the stars in the field are monitored continuously regardless of whether 
they are being lensed or not, planets can be detected at very large 
projected separations when the primary star is not significantly 
magnifying the source, and indeed even when the signature of the 
primary is absent.  Therefore these surveys are expected to be 
sensitive to both wide-separation \citep{distefano99a, distefano99b} 
and even free-floating \citep{bennett02, han03} planets.  Such planet
populations are difficult or impossible to probe by other planet
search techniques.  Several such high-frequency experiments in space
and on the ground have already been proposed or are being seriously
considered.  The proposed space microlensing missions of {\it Galactic
Exoplanet Survey Telescope} ({\it GEST}, \citealt{bennett02}) and {\it
Joint Dark Energy Exoplanet Mission} ({\it JDEEM}) are designed to
continuously monitor $\sim 10^8$ Galactic bulge main-sequence stars
with $\sim 1\%$ photometric precision and a frequency of several times
per hour by using a 1--2 m aperture space telescope.  Detailed
simulations of the outcomes of a ground-based high-frequency
experiment using a network of 2~m-class telescopes are being
carried out by \citet{gaudi04}.

Although efficient in detecting planets, the microlensing method 
suffers the shortcoming that the mass of the detected planet is 
generally poorly constrained.  Analysis of the planet-induced 
perturbation automatically yields the planet-star mass ratio $q=m_p/M$, 
but the stellar mass is unknown due to the degeneracy of the physical 
lens parameters in the principal lensing observable.  This degeneracy 
arises because among the three observables related to the physical 
parameters of the lens, the Einstein timescale $\te$, the angular 
Einstein radius $\thetae$, and the Einstein ring radius projected 
onto the observer plane $\retilde$, only $\te$ is routinely measurable 
from the lensing light curve.  These three observables are related to 
the underlying physical lens parameters of the mass $M$, relative 
lens-source parallax $\pi_{\rm rel}={\rm AU}\ (D_{ol}^{-1}-D_{os}^{-1})$, 
and proper motion $\mu_{\rm rel}$, by
\begin{equation}
\te = {\thetae\over \mu_{\rm rel}},\ \ \ 
\thetae=\sqrt{4GM\pi_{\rm rel} \over c^2\ {\rm AU}},\ \ \ 
\retilde = \sqrt{4GM\ {\rm AU} \over c^2 \pi_{\rm rel}}.
\label{eq2}
\end{equation}
For the unique determination of the planet mass, one must, therefore, 
measure the other two observables of $\thetae$ and $\retilde$.  Once 
$\thetae$ and $\retilde$ are known, the lens mass is determined by
\begin{equation}
M=\left( {c^2 \over 4G} \right)\retilde \thetae.
\label{eq3}
\end{equation}

There have been several methods proposed to determine the masses of
planets detectable by future lensing experiments. \citet{bennett02}
pointed out that for some events with detected planets, the proposed
space lensing mission would detect enough light from the lens star to
determine its spectral type and so infer the mass.  However, no more
than $\sim 1/3$ of lenses are bright enough to be so detected.  A more
direct method of determining masses of bound planets was proposed by
\citet{gould03}.  They first demonstrated that the precision and sampling 
of the space observations will be sufficient to routinely detect one 
projection of the vector quantity\footnote{The vector $\pivec_{\rm E}
={\rm AU}/\tilde{\rvec}_{\rm E}$ has the magnitude of $\retilde$ and 
the direction of relative source-lens proper motion.} $\pivec_{\rm E}$
from the primary event.  They then demonstrated that a second
projection of $\pivec_{\rm E}$ could be measured by combining
observations from a satellite in an L2 orbit with ground-based
observations. For this setup, an Earth-satellite baseline of $d_{sat}
\sim 0.005\ {\rm AU}$ is sufficient to routinely detect the difference
in the peak time of the planetary perturbation as seen from the Earth
and satellite for low-mass planets.  The difference can then be
combined with the known Earth-satellite projected separation to
measure the second projection of $\pivec_{\rm E}$.  The two projections 
of $\pivec_{\rm E}$ yield the magnitude of $\retilde$.  Moreover, for 
terrestrial planets with mass ratio $q\sim 10^{-5}$, the Einstein ring 
radii $\thetaep=q^{1/2}\thetae$ is of order the angular source size 
$\theta_\ast$ of a typical (main-sequence) source.  Therefore, the 
magnification pattern arising from the planet typically has structure 
on the scale of the source size, which gives rise to finite-source 
effects on the perturbation lightcurve.  From these effects, it would 
be possible to measure $\rho_{\ast}=\theta_\ast/\thetae$.  The angular 
size $\theta_\ast$ can be determined from its dereddened color and 
magnitude using an empirically-calibrated color-surface brightness 
relation.  Thus a measurement of $\rho_{\ast}$ can be used to infer 
$\thetae$ and thus complete the mass measurement via equation~(\ref{eq3}).

\citet{han04} demonstrated that the same observational setup discussed
by \citet{gould03} to determine the mass of bound planets can also be
used to determine the masses of free-floating planets.  The principles
are generally the same, however the primary difference is the lack of
structure in the planet deviation caused by the small caustic residing 
at the center of the planet's Einstein ring that exists for bound planets.  
This lack of structure is beneficial because it allows one to unambiguously
determine both components of $\pivec_{\rm E}$ from the planetary event 
itself.  However, it also implies that the cross section for significant 
finite-source effects is much smaller than in the bound-planet case, 
for which the caustic structure is significantly extended.  Essentially, 
it is only possible to detect finite-source effects, and so measure 
$\thetae$, for those events in which the source star transits the planet.  
The angular radius $\theta_\ast$ of a typical main-sequence source star 
in units of the Einstein ring radius $\theta_{{\rm E},p}$ of a planet is,
\begin{equation}
\rho_{\ast,p} \equiv {\theta_\ast \over \thetaep}\sim 0.6\left({m_p
\over M_{\oplus}}\right)^{-1/2},
\label{eq4}
\end{equation}
where $m_p$ is the mass of the planet, and we have assumed 
$\theta_*\simeq 0.6~{\mu{\rm as}}$, which is typical for main-sequence 
sources in the bulge, and $\pi_{\rm rel}\simeq 42~{\mu{\rm as}}$ (i.e.\ 
$D_{ol}=6~{\rm kpc}$ and $D_{os}=8~{\rm kpc}$).  Therefore, mass 
measurements are only routine for planets with masses for which 
$\rho_{\ast,p}\ga 1$, i.e., essentially $m_p\la 0.3~M_\oplus$, and are 
less common for larger mass planets.

Events caused by wide-separation planets have similarities to events
produced by both free-floating planets as well as events caused by
planets in the lensing zone.  The gross structure of events caused by
wide-separation planets is similar to that of free-floating planets,
and thus the mass of the planet can be determined in a similar way.
For wide-separation planetary events, however, the primary (parent star) 
provides a shear at the location of the planet of $\gamma = s^{-2}$.
This shear produces a caustic of angular width $\sim 4\gamma \thetaep$
near the location of the planet.  This small caustic can cause anomalies 
near the peak of the light curves. \citet{han03} pointed out that these 
anomalies can be used to distinguish events caused by bound planets
from those caused by free-floating planets.  In addition, this extended 
caustic structure increases the cross-section for significant finite 
source-size effects, thereby increasing the fraction of events for 
which it is possible to measure $\thetaep$ (and thus the planet mass) 
relative to the unbound case.

In this paper, we present a comprehensive discussion of the ability of
microlensing to characterize wide-separation planets, consolidating
and augmenting the studies of \citet{gould03}, \citet{han04}, and
\citet{han03}.  We consider the ability of microlensing to distinguish
free-floating planets from bound planets, as well to measure the mass
and projected physical separation of wide-separation planets, as a
function of these parameters.  The paper is organized as follows.  In
\S\ 2, we discuss the lensing characteristics of wide-separation
planetary events.  In \S\ 3, we estimate the probability of measuring
$\rho_{\ast,p}$ from future space observations of wide-separation
planetary events by performing detailed simulations of these events,
and assessing the resulting uncertainties in $\rho_{\ast,p}$.  In \S\
4, we summarize the ability of microlensing to characterize
wide-separation planets, as well as discuss methods of distinguishing
planetary lensing events caused by free-floating planets from those
caused by wide-separation planets.  We conclude in \S\ 5.

Throughout this paper, we assume that the lens system (star, planet,
or both) is located at $D_{ol}=6~{\rm kpc}$, and that the source is a
solar-type star with radius $R=R_\odot$ located at $D_{os}=8~{\rm kpc}$.  
Thus
$\theta_*=0.58~\mu{\rm as}$ and $\pi_{\rm rel}=41.7~{\mu{\rm as}}$.
The relative proper motion is assumed to be $\mu_{\rm rel}=26.0~{\rm
km~s^{-1}~kpc^{-1}}$, which yields, for a (star or planet) lens of
mass $M$, an event timescale of $\te\simeq 38.8~{\rm days}
(M/M_\odot)^{1/2}$, an angular Einstein ring radius of $\thetae \simeq
582~{\mu{\rm as}}(M/M_\odot)^{1/2}$, a projected Einstein ring radius
of $\retilde\simeq 13.9~{\rm AU}(M/M_\odot)^{1/2}$, and a dimensionless 
source size of $\rho_\ast \simeq 10^{-3} (M/M_\odot)^{1/2}$.  When 
specified, we assume a primary mass of $M=0.3~M_\odot$.

\begin{figure}[t]
\epsscale{1.2}
\plotone{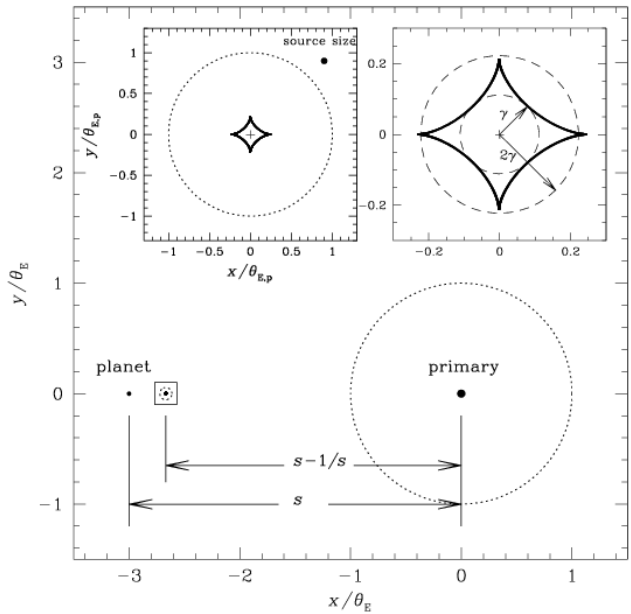}
\vskip-0.2cm
\caption{\label{fig:one}
Lensing geometry of a star with a wide-separation planet.  Shown are 
the locations of the star (at the origin), planet (at $[x,y]=[-3,0]\thetae$), 
and the resulting caustics. The tiny central caustic is located very 
close to the star, and the planetary caustic is located in the region 
enclosed by a small box. The upper left inset is a blowup of the region 
around the planetary caustic.  The dotted circles represent the Einstein 
rings of the individual lens components around their effective lens 
positions.  The upper right inset shows the planetary caustic enclosed 
by two circles with radii of $1.0\gamma$ and $2.0\gamma$, respectively, 
where $\gamma$ is the shear.  The geometry is for the case in which the 
planet/star mass ratio is $q=3\times 10^{-3}$ and the normalized separation 
is $s=3$.  The small disk at the upper right corner of the left inset 
represents the size of a source star with a radius $R=1.0\ R_\odot$ at 
$D_{os}=8$ kpc.  For the lens, we assume $M=0.3~M_\odot$ and $D_{ol}=6$ 
kpc, and thus the angular Einstein radius is $\thetae=0.32~{\rm mas}$.
Note that the axes of the main panel are scaled by the combined Einstein 
radius ($\thetae$), while the axes of the insets are scaled by the 
Einstein ring radius of the planet ($\theta_{{\rm E},p}=\sqrt{q}\thetae$).
}\end{figure}

\section{Wide-Separation Planetary Events}

Generally, one can write the mapping from the lens plane to the source 
plane of $N$ point masses with no external shear or convergence as,
\begin{equation}
\zeta = z - \sum_{j=1}^N {m_j/M \over \bar{z}-\bar{z}_{L,j}},
\label{eq5}
\end{equation}
\citep{witt90}, where $\zeta=\xi + i\eta$, $z_{L,j}=x_{L,j}+iy_{L,j}$, 
and $z=x+iy$ denote the source, lens, and image positions, respectively, 
$\bar{z}$ denotes the complex conjugate of $z$, and $m_j/M$ are the mass
fractions of the individual lens components ($\sum_j m_j=M$).  Here
all angles are normalized to the Einstein ring radius $\thetae$ of the
total mass of the system $M$.  The lensing process conserves the
source surface brightness, and thus the magnifications $A_i$ of the
individual images $i$ correspond to the ratios between the areas of
the images and the source.  For an infinitesimally small source
element, this is,
\begin{equation}
A_i = \left\vert \left( 1-{\partial\zeta\over\partial\bar{z}}
{\overline{\partial\zeta}\over\partial\bar{z}} \right)^{-1}
\right\vert.
\label{eq6}
\end{equation}
The total magnification is just the sum over all images, $A=\sum_i A_i$.

For a single lens ($N=1$), one can easily invert the lens equation to
solve for image positions $(x,y)$ and magnifications as a function of 
the source position $(\xi,\eta)$.  This yields the familiar result that
there are two images for every source position $\zeta \ne 0$.  These
two images have angular separations $\theta \equiv |z-z_{L}|$ from
the lens of $\theta_{\pm} = 0.5[u \pm (u^2 +4)^{1/2}]$, where $u\equiv
|\zeta-z_{L}|$.  The image $\theta_+ > 1$ ($\theta_- <1$) is often 
referred to as the major (minor) image

A planetary lens is described by the formalism of a binary ($N=2$) 
lens.  In this case, the lens equation cannot b inverted algebraically.  
However, it can be expressed as a fifth-order polynomial in $z$ and 
the image positions are then obtained by numerically solving the 
polynomial \citep{wm95}.  One important characteristic of binary 
lensing is the formation of caustics, which represent the set of 
source positions at which the magnification of a point source becomes 
infinite.  The number and size of these caustics depends on the 
projected separation $s$ and the mass ratio $q$.

One can think of a wide-separation planet with $s\gg 1$ and $q \ll 1$
as a perturbation to the major image produced by the primary.  The
location of the major image produced by the primary is $\theta_+=
0.5[u + (u^2 +4)^{1/2}]$, and therefore a planet separated by $s$ 
from its parent star will produce a `planetary' caustic on the 
star-planet axis at an angular separation from its parent star of 
${\hat s}=s-1/s$ (see Fig.~\ref{fig:one}).  In addition, there will 
be a second, smaller `central' caustic located near the star 
\citep{griest98}.  By choosing the origin in the image plane as the 
position of the planet, the origin in the source plane as the point 
on the star-planet axis with an angular separation ${\hat s}$ from 
the primary, and then normalizing all angles to the Einstein ring of 
the secondary, it is straightforward to show that, in the limit $s\gg1$ 
and $q\ll1$, the binary-lens equation becomes
\begin{equation}
\hat{\zeta} = \hat{z} - {1 \over \hat{\bar{z}}} +\gamma \hat{\bar{z}},
\label{eq7}
\end{equation}
\citep{dominik99}. Here $\gamma=s^{-2}$ is the shear and the notations 
with upper ``hat'' mark represent the length scales normalized by the 
Einstein radius corresponding to $m_2$, e.g.\ $\hat{z}=z (\thetae/ 
\theta_{\rm E,2})$.  This is the well-known \citet{chang79, chang84} 
lens.  In its range of validity, equation~(\ref{eq7}) implies that if 
the planetary separation is sufficiently wide, the lensing behavior 
in the region around the planetary Einstein ring can be approximated 
as a single point-mass lens superimposed on a uniform background shear 
$\gamma$.  The caustics created by a Chang-Refsdal lens with $\gamma<1$ 
have an asteroid shape (see Fig.~\ref{fig:one}) with full width along 
the star-planet axis and height normal to the planet-star axis of, 
respectively,
\begin{equation}
2a_{\rm C-R} = {4\gamma\over {\sqrt{1-\gamma}}}, \qquad 2b_{\rm C-R} =
{4\gamma\over {\sqrt{1+\gamma}}}.
\label{eq8}
\end{equation}
Thus, as the separation between the star and planet increases, the
size of the caustic shrinks approximately as $1/s^2$, and both
lens components tend to behave as if they are two independent single
lenses.

\begin{figure}[t]
\vskip-0.2cm
\epsscale{1.2}
\plotone{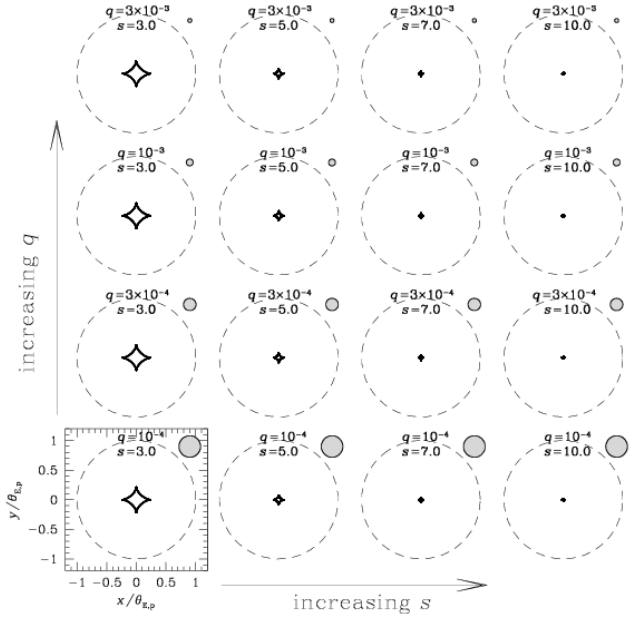}
\caption{\label{fig:two}
Variations of the caustic and source sizes relative to the size of the 
planetary Einstein ring of wide-separation planets with various mass 
ratios relative to and separations from the primary star.  In each panel, 
the diamond-shaped figure is the planetary caustic, the dashed circle 
represents the Einstein ring of the planet, and the shaded small 
circle in the upper right corner represents the size of the source 
star relative to the planetary Einstein ring.  Panels are arranged 
so that the separation and mass ratio are increasing along the 
abscissa and ordinate, respectively. The assumed lens geometry is 
that the lens and source star are located at $D_{ol}=6$ kpc and 
$D_{os}=8$ kpc, respectively, and the source star has a physical 
radius of $R_\ast=1.0 R_\odot$.
}\end{figure}

For the applications discussed here, the angular size of the source 
is typically non-negligible relative to the Einstein ring of the
secondary, $\theta_{{\rm E},p}$.  Therefore, finite source effects
must be taken into account.  It is precisely these finite source
effects that allows one to measure $\thetaep$.  The magnification of 
a finite source is computed by the intensity-weighted magnification
averaged over the source star flux, i.e.,
\begin{equation}
A_{fs}(\zetavec)=
{\int_S I(\zetavec') A(\zetavec+\zetavec') d \zetavec'
\over 
\int_S I(\zetavec')d \zetavec'},
\label{eq9}
\end{equation}
where $A$ denotes the magnification of the corresponding 
point source, $\zetavec$ is the vector position of the center of 
the source, $\zetavec'$ is the displacement vector of a point on 
the source star surface with respect to the source star's center, 
and the two-dimensional integral is over the source-star surface $S$.  
For a source with uniform surface brightness, the computation can be 
reduced from a two-dimensional to a one-dimensional integral using 
the Generalized Stokes's theorem \citep{gould97, dominik98}.  The 
finite-source effect becomes important when the source passes over 
a magnification pattern with small-scale structure or crosses the 
caustics. Because wide-separation planetary events have such extended
magnification structures (i.e., the planetary caustics), the chance 
to measure $\rho_{\ast,p}$ is expected to be higher compared to the 
case of free-floating planet events.

In Figure~\ref{fig:two}, we show how the relative caustic and source
sizes vary with respect to the Einstein radius for wide-separation
planets with various mass ratios relative to and separations from the
primary star, using the full binary-lens formalism.  For a fixed
angular source size $\theta_*$, the normalized source size
$\rho_{\ast,p}$ becomes smaller with increasing planet/star mass ratio
(as $\rho_{\ast,p}\propto q^{-1/2}$).  However, in the parameter
regimes shown in the figure, the Chang-Refsdal approximation is
excellent, and therefore the size of the caustic in units of $\thetaep$ 
depends almost solely on $s$.  Thus, as the separation increases, the 
caustic width decreases as $\sim 4\gamma \propto s^{-2}$.

\section{Probability of Measuring $\thetaep$}

We now address the question of what fraction of wide-separation
planetary events in future lensing experiments will yield an accurate
measurement of $\thetaep$. We estimate the probability, $P$, that a 
given wide-separation planetary event will exhibit substantial finite 
source effects and so allow the measurement of $\rho_{\ast,p}$ to a 
given accuracy (in turn yielding $\thetaep$).  In particular, we are
interested in how the existence of the caustic created by the shear
from the primary increases the probability relative to a free-floating
planet event with similar lensing characteristics.

In order to estimate the probability $P$, we carry out detailed 
simulations of wide-separation planetary events and estimate the 
uncertainties of $\rho_{\ast,p}$ determined from simulated light curves.
Although wide-separation planetary events are reasonably well-described
by the Chang-Refsdal approximation, in order to make our results fully 
general, we will carry out our simulations using the full binary-lens 
formalism.  However, we will use the Chang-Refsdal approximation to aid 
in the interpretation of our results.  The simulations proceed as follows.  
Planetary microlensing event lightcurves are calculated using equations 
(\ref{eq5}) and (\ref{eq6}).  The ranges of the planetary separations 
and mass ratios of the tested events are $2\leq s \leq 20$ and $10^{-5}
\leq q \leq 10^{-2}$, respectively.  Finite-source effects are incorporated
by computing a one-dimensional line integral along the boundaries of 
the images, whose positions are obtained by numerically solving the 
lens equation, and then applying Stokes's theorem.  The majority of 
target stars to be monitored by the proposed space microlensing mission 
are Galactic bulge main-sequence stars, we therefore assume a solar-type 
source star with an apparent magnitude of $I\sim 21$.  For the space 
observations, we follow the specification of the {\it GEST} mission 
\citep{bennett02} and assume that events are monitored with a frequency 
of $f_{obs}=5~{\rm hr^{-1}}$ and that a 600-second exposure image is 
acquired from each observation.  We assume a photon acquisition rate of 
13 photons/s for an $I=22$ star.

Once events are produced from the simulation, the uncertainties of the 
individual fitting parameters ($p_i$) are determined from each light 
curve by
\begin{equation}
\sigma_{i}=\sqrt{c_{ii}};\qquad c=b^{-1},
\label{eq10}
\end{equation}
where $c_{ij}$ is the covariance matrix, and the curvature matrix of 
the $\chi^2$ surface is defined by
\begin{equation}
b_{ij}=\sum_{k=1}^{N_{\rm obs}} {\partial F_k\over \partial p_i}
{\partial F_k\over \partial p_j} {1\over \sigma_k^2 }.
\label{eq11}
\end{equation}
Here $F_k(t) = A(t_k)F_S+F_B$ represents the observed flux for each 
measurement, $F_S$ and $F_B$ are the fluxes of the source and blended 
light, $N_{\rm obs}$ is the total number of measurements, and $\sigma_k$ 
is the measurement error.  For planetary lensing, the total number of 
parameters is 9 including $s$, $q$, $F_S$, $F_B$, $\rho_{\ast,p}$, $\tep$, 
$t_0$, $\alpha$, and $u_{0,p}$, where we define $\tep$ as the time required 
for the source to transit $\theta_{{\rm E},p}$, $t_0$ is the time of 
the closest approach to the planetary caustic, $u_{0,p}$ is the separation 
(normalized by $\theta_{{\rm E},p}$) at that moment, and $\alpha$ is 
the orientation angle of the source trajectory with respect to the 
star-planet axis.  According to the luminosity function of the Galactic 
bulge field \citep{holtzman98}, the surface number density of stars 
with $I\la 24$ is $\sim 5,000\ {\rm stars}/{\rm arcmin}^2 \sim 1.4\ 
{\rm stars}/ {\rm arcsec}^2$, and thus a space mission equipped with
a $\sim 1$ m telescope can resolve most neighboring stars.  However, since 
one cannot exclude the possibility of blending by the light from a 
companion to the source or by the lens itself, we include the blending 
parameter $F_B$.  For the simulation, we set the blending fraction to 
be $F_B/F_S=0.3$.  The photometric uncertainty is assumed to be limited 
by photon statistics and the uncertainties of the fitting parameters 
are determined from the light curve measured during 
$-2.0~\tep \le t-t_0 \le 2.0~\tep$.

\begin{figure}[t]
\vskip-0.4cm
\epsscale{1.2}
\plotone{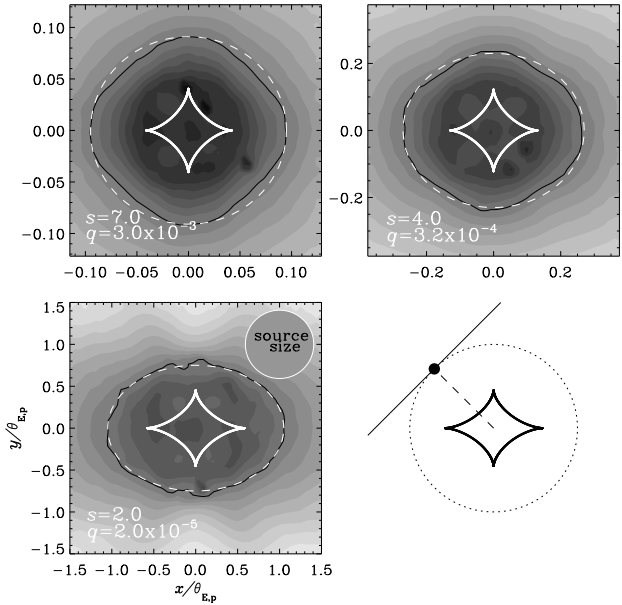}
\caption{\label{fig:three}
Contour maps of the fractional uncertainty 
$\sigma_{\rho_{\ast,p}}/\rho_{\ast,p}$ as a function of the source 
position relative to the planetary caustic.  Each position on the map 
represents a source trajectory that passes through that position and 
is normal to the line connecting the position and the caustic center 
(see the geometry in the lower right panel).  For all three presented 
cases, the source size relative to the caustic size is nearly the same.  
The dark closed curve in each panel represents the contour at which 
$\sigma_{\rho_{\ast,p}}/\rho_{\ast,p}=10\%$ and the white dashed curve 
is the approximation of this contour as an ellipse.
}\end{figure}

As we argued in \S\ 2, the lensing behavior of a wide-separation
planet is locally well-described by a single lens with external shear
$\gamma=s^{-2}$, i.e.\ a Chang-Refsdal lens.  Therefore, the
magnification structure can be described by only two parameters,
namely $(\gamma,\rho_{\ast,p})$, rather than the three parameters
$(s,q,\rho_{\ast})$ generically required for full binary lensing.
In fact, for $\gamma \ll 1$ ($s \gg 1$), the Chang-Refsdal lens
becomes self-similar: when all angles are normalized to $2\gamma$, 
the magnification pattern is nearly independent of $\gamma$.  
Therefore, the ability to measure $\rho_{\ast,p}$ depends primarily
on the single parameter $\rho_{\ast,p}/\gamma$.  In fact, this
scaling breaks down for two reasons.  First, as $\gamma$ approaches
unity, the self-similarity of the Chang-Refsdal lens breaks down. 
Second, because we are considering a fixed sampling rate, the number
of samples per planetary crossing time $t_{{\rm E},p}=q^{1/2}\te$
decreases for smaller-mass planets.  Therefore, the error on the light
curve fit parameters, which depends on both the magnification
structure and the density of data points, formally depends on $q$ as
well.  However, these two effects are generally sub-dominant,
and therefore the dependence on $\gamma$ or $q$ for fixed 
$\rho_{\ast,p}/\gamma$ is relatively weak.  We
demonstrate this by computing the fractional uncertainty
$\sigma_{\rho_{\ast,p}}/\rho_{\ast,p}$ as a function of the source
positions around the planetary caustic.  Figure~\ref{fig:three} shows
the contour maps of $\sigma_{\rho_{\ast,p}}/\rho_{\ast,p}$ for three
different combinations of $s$ and $q$ yielding the same ratio of
$\gamma/\rho_{\ast,p}$.  In the maps, each position represents a
source trajectory that passes through that position and is normal to
the line connecting the position and the caustic center (see the
geometry illustrated in the lower right panel), and the closed curve
drawn by a dark line represents the contour at which
$\sigma_{\rho_{\ast,p}}/\rho_{\ast,p}=10\%$.  From a comparison of the
maps, one finds that the patterns look qualitatively similar despite
the great differences in the values of $s$ and $q$, confirming the
argument that the ability to measure $\rho_{\ast,p}$ depends
primarily on $\rho_{\ast,p}/\gamma$.

\begin{figure}[t]
\epsscale{1.2}
\plotone{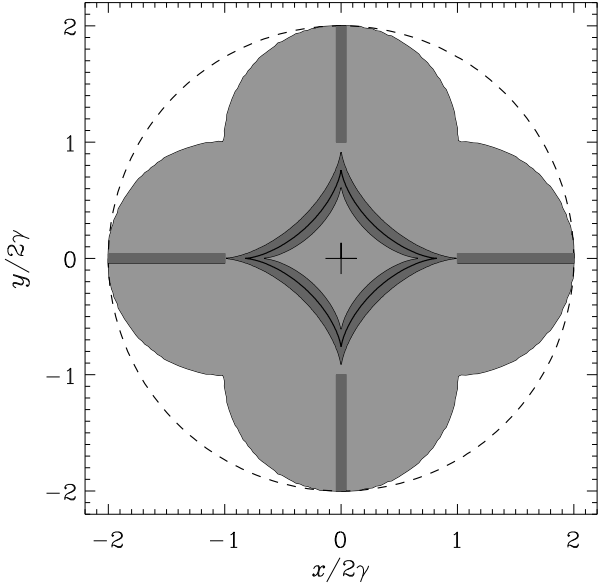}
\caption{\label{fig:four}
Cartoon showing the validity of approximating the effective region for 
$\rho_{\ast,p}$ measurement as an ellipse.  Here the dark shaded area 
represents the source positions where, for $\rho_{\ast,p}\ll \gamma$, 
finite source effects are large, thus allowing $\rho_{\ast,p}$ to be 
effectively measured.  The light shaded region enclosed by the 
clover-shaped figure represent the effective region of source 
trajectories that pass the dark shaded region, and the dashed 
circle represents the boundary of the effective region following the 
ellipse approximation.  We note that the cross sections of the dark 
and light shaded regions are, by definition, same.  All lengths are 
normalized by the caustic size, i.e., $2\gamma$.
}\end{figure}

A straightforward approach to estimate $P$ would be first to construct
maps of $\sigma_{\rho_{\ast,p}}/\rho_{\ast,p}$ such as the ones shown
in Figure~\ref{fig:three} for the various combinations of $s$ and $q$,
second to draw many light curves with various combinations of $u_{0,p}$
and $\alpha$ obtained from one-dimensional cuts through each constructed 
map, and then to estimate the probability as the ratio between the number 
of light curves yielding uncertainties smaller than a threshold value 
out of the total number of tested events.  However, we find that this 
approach is difficult to implement because constructing the large number 
of high-resolution maps that incorporate finite-source effects demands 
a large amount of computation time even after the great reduction of 
finite-source calculations using Stokes's theorem (i.e., from a 
two-dimensional to a one-dimensional integral).  Fortunately, we find 
that the region where $\rho_{\ast,p}$ can be measured to a given precision 
(the effective region) is well confined around the planetary caustic and 
its boundary is, in general, approximated as an ellipse as illustrated 
in Figure~\ref{fig:three}.  We therefore estimate $P$ by determining 
the semi-major, $a$, and semi-minor, $b$, axes of the ellipse and then 
compute the probability, which corresponds to the ratio of the 
angle-averaged cross-section of the ellipse to the diameter of the 
planetary Einstein ring\footnote{The approximation of the boundary of 
the effective region as an ellipse becomes poor as the source size 
becomes smaller than the caustic size.  In the limiting case 
$\rho_{\ast,p}\ll \gamma$, the positions at which finite source effects 
are large (allowing $\rho_{\ast,p}$ to be effectively measured) are 
confined to regions near the caustic itself, as well as the protruding 
region outside of the caustic cusps (the dark shaded regions in 
Fig.~\ref{fig:four}).  Even in this limiting case, however, we note that 
the probability $P$ determined by eq.~(\ref{eq12}) is still a good 
approximation.  This is demonstrated in Fig.~\ref{fig:four}, in which 
the light shaded region enclosed by the clover-shaped figure represents 
the effective region of source trajectories that can pass the dark shaded
region, and the dashed circle represents the boundary of the effective
region following the ellipse approximation.  We find that even in this
extreme case, the ratio between the angle-averaged cross-sections of
the clover-shaped and circular regions is $0.903$, implying that the
error in the probability as determined by the ellipse approximation
is not important.}, by
\begin{equation}
P={1\over \pi}
\int_0^\pi \sqrt{a^2\sin^2 \alpha + b^2\cos^2 \alpha}\ d\alpha
={2\over \pi} a E(e),
\label{eq12}
\end{equation}
where $E$ represents the complete elliptical integral of the second 
kind and $e=(1-b^2/a^2)^{1/2}$.  Note that, since $a$ and $b$ are in 
units of $\theta_{{\rm E},p}$, the probability is normalized such that 
$P$ is the fraction of events with $u_{0,p}\le 1$ that yield a measurement 
of $\rho_{\ast,p}$ to a given factional precision 
$\sigma_{\rho_{\ast,p}}/\rho_{\ast,p}$.  Planetary events with $u_{0,p}>1$ 
may well be detectable in the next-generation lensing surveys, and thus 
the fraction of detectable planetary lensing events with a measurement
of $\rho_{\ast,p}$ to a given accuracy is likely to be smaller.  We 
address this point in \S\ 4.

\begin{figure}[t]
\epsscale{1.15}
\plotone{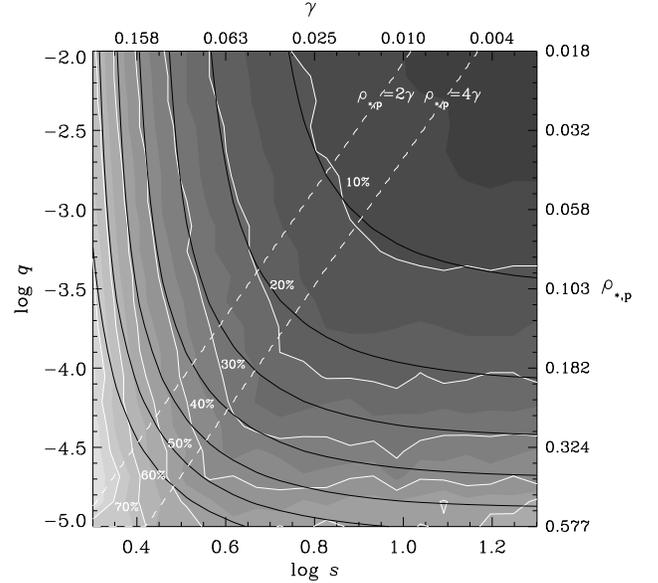}
\caption{\label{fig:five}
Probability $P$ of measuring $\rho_{\ast,p}$ for wide-separation 
planetary events as a function of the planetary separation and mass 
ratio.  The adopted threshold uncertainty is $(\sigma_{\rho_{\ast,p}}
/\rho_{\ast,p})_{th}=10\%$.  The two sets of contours drawn by dark 
and white lines are based on numerical computations considering the 
full binary lensing and an analytic treatment using the Chang-Refsdal 
lensing, respectively (see the text for details).  The grey-scale tones 
show the probability $P$ based on numerical computations: they change 
for every 5\% change of $P$.  The two dashed straight lines represent 
the positions at which $\rho_{\ast,p}=2.0\gamma$ and $4.0\gamma$, 
respectively.
}\end{figure}

Figure~\ref{fig:five} shows the contour map (marked by white contours 
and grey scales) of the determined probability $P$ as a function of 
the planetary separation and mass ratio.  For the map, the imposed 
threshold uncertainty is $(\sigma_{\rho_{\ast,p}}/\rho_{\ast,p})_{th}
=10\%$.  From the probability map, we find two basic regimes.
First, in the region of the parameter space where the 
caustic size is much smaller than the source size, the 
contribution of the caustic to the probability is not important and 
the value of $P$ is essentially the same as that of the corresponding 
free-floating events.  Since $\rho_{\ast,p}$, which is the dominant 
factor determining $P$ in this region, depends only on the mass ratio 
and does not depend the separation, the probability distribution is 
flat as shown in the right lower part of Figure~\ref{fig:five}. On 
the other hand, in the region where the caustic size is of the order 
or larger than the source size, the contribution of the caustic to 
$P$ becomes important.  The caustic size, and thus $P$, rapidly 
increase with the decrease of the planetary separation, and this is 
reflected in the probability distribution trends in the upper left 
part of Figure~\ref{fig:five}.  We find that the two regions (i.e., 
source- and caustic-size dominant regions) are divided roughly by the 
line $\rho_{\ast,p} = 4\gamma$.  Knowing that the relative cross-section 
for $\rho_{\ast,p}$ measurement results from the combination of the 
caustic and normalized source sizes, and keeping the validity of the 
Chang-Refsdal lensing in mind, we interpret the result by analytically 
assessing the probability based on the semi-major and semi-minor axes 
of the effective region as the linear combination of the Chang-Refsdal 
caustic size and the normalized source radius, i.e., 
\begin{equation}
\eqalign{
a={\cal C}_\gamma a_{\rm C-R}+{\cal C}_{\rho_\ast} \rho_{\ast,p}, \cr
b={\cal C}_\gamma b_{\rm C-R}+{\cal C}_{\rho_\ast} \rho_{\ast,p}, \cr
}
\label{eq13}
\end{equation}
where $a_{\rm C-R}$ and $b_{\rm C-R}$ are defined in
equation~(\ref{eq8}), and ${\cal C}_\gamma$ and ${\cal
C}_{\rho_\ast}$ are linear coefficients.  In the point-mass limit
($\gamma \ll \rho_{\ast,p}$), it is known that $\rho_{\ast,p}$ can be
measured only when the lens crosses the source star \citep{gould96},
and thus we set ${\cal C}_{\rho_\ast}=1.0$.  By adjusting ${\cal
C}_\gamma$, we find that ${\cal C}_\gamma\sim 2.5$ yields the best-fit
probability distribution (marked by dark contours in
Fig.~\ref{fig:five}) to the one based on numerical computations.
This implies that in the caustic-dominant regime, the effective
cross section of the caustic for a $\rho_{\ast,p}$ measurement is larger
than its full width by $\sim 25\%$, i.e.\ $a/(4\gamma)\simeq 1.25$.

Translating a measurement of $\rho_{\ast,p}$ into a measurement of
$\thetaep$ requires knowledge of the angular size of the source
$\theta_\ast$.  This can be estimated from the known source color and
magnitude as described in \citet{yoo04a}.  Briefly, the process works
as follows.  The apparent source color and apparent magnitude can be 
estimated from multicolor photometry taken at several different source 
magnifications.  The dereddenned color and magnitude can then be found 
by comparing to the apparent color and magnitude of nearby stars in
the red clump, whose dereddened color and magnitude are known.  
This assumes that the source star is being seen through the same column
of dust as the stars in the red clump.  In practice, however, even 
fairly large differences in the dust column have relatively little effect 
because the source color and magnitude have opposite effects on the 
inferred value of $\theta_\ast$.  The angular size $\theta_*$ of the 
source can then be inferred from is color and magnitude using an 
empirical color-surface brightness relation (e.g.\ \citealt{vb99}).  
The statistical error in the derived value of $\theta_*$ from this 
procedure will likely be dominated by the intrinsic scatter in the 
empirical color-surface brightness relation, which is currently 
$\sim 10\%$.  Therefore, the error in the inferred value of 
$\thetaep$ will be dominated by the error in $\rho_{\ast,p}$ for 
$\sigma_{\rho_{\ast,p}}/\rho_{\ast,p}\ga 10\%$.

\section{Characterization of Wide-Separation Planets}

In this section, we summarize the potential of next-generation
microlensing surveys to characterize wide-separation planets, focusing
on their ability to distinguish isolated planetary events caused by
free-floating planets from those caused by wide-separation planets
as a function of the separation of the planets from their host stars.
As the formation processes and evolution histories of wide-separation
and free-floating planets are believed to be substantially different,
unless these two populations of planetary events can be distinguished,
it will be difficult to extract useful information about the formation
and evolution of these individual planet populations.\footnote{In fact, 
the frequency of bound and free-floating planets must be determined 
statistically from the ensemble of observed planetary events.  As we 
demonstrate in \S 4.3, the majority of events from wide planets
with $r_\perp \ga 20~{\rm AU}$ will show no signature of the primary,
and so it is not generically possible to distinguish between bound and
free-floating planets on an event-by-event basis.  Therefore, it will 
be necessary to use those events that are known to be due to bound
planets to statistically infer the fraction of events with no signature 
of a primary that are due to bound planets.} We also summarize the 
capacity of these surveys to measure the masses of the planets through 
the measurement of $\thetae$ and $\retilde$, as a function of the planet 
mass and separation.   The majority of the calculations in the following 
sections are independent of the mass of the primary, and so our results 
are most naturally expressed in terms of their dependence on the planet 
mass $m_p$, rather than the planet mass ratio, $q$.  However, to make 
contact with the results from the previous sections, we will also quote 
results in terms of $q$, assuming a primary mass of $M=0.3~M_\odot$, 
i.e.\ $q=10^{-5}(m_p/M_\oplus)$.

\begin{figure}[t]
\epsscale{1.15}
\plotone{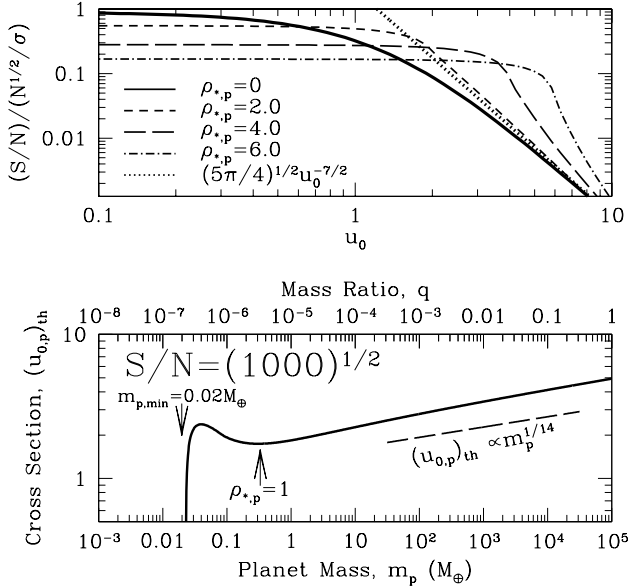}
\caption{\label{fig:six}
Top panel: The curves show the normalized signal-to-noise ratio
$(\sn)/(N^{1/2}/\sigma)$, as a function of the impact parameter 
$u_0$, for single lens events with various values of the source 
size $\rho_{\ast,p}$ in units of the Einstein ring radius.  Here 
$N$ is the number of points per Einstein crossing time each with 
fractional precision $\sigma$. The solid curve is for a point source
($\rho_{\ast,p}=0$), and the dotted curve shows the limiting form for
a point source light curve with $u_0\gg 1$, i.e.\ eq.~(\ref{eq15}).
Bottom panel: The solid curve shows the cross-section $(u_{0,p})_{th}$ to
detect a planet at $\sn \ge (1000)^{1/2}$ as a function of its mass $m_p$
 (bottom axis) and mass ratio $q$ (top axis).  The long-dashed line shows 
the scaling $(u_{0,p})_{th}\propto m_p^{1/14}$, which is valid for point 
sources and $(u_{0,p})_{th}\gg 1$.  Also shown are the minimum detectable 
planet mass, $m_{p,min}\simeq 0.02~M_\oplus$, and the mass at which 
$\rho_{\ast,p}=1$.
}
\end{figure}

\subsection{Cross Section for Detection}

We first address the question of the detectability of planets as a
function of their mass.  For fixed $\mu_{\rm rel}$ and $\pi_{\rm rel}$, 
and for identical observational setups, larger-mass planets are detectable 
to larger planetary impact parameters $u_{0,p}$ because the number of 
points per crossing time is larger, thereby increasing the $\sn$.  
This implies a larger cross section for detection.  As the mass ratio 
decreases, the minimum impact parameter required to produce a lightcurve 
above a given $\sn$ decreases, thus decreasing the cross-section for 
detection.  Eventually, finite source size effects become important 
when $\rho_{\ast,p}\ga 1$.  For sufficiently high photometric accuracy, 
finite source effects can increase the cross section, since the 
timescale for the event will be set by the crossing time of the source, 
$t_\ast\equiv \rho_{\ast}/\mu_{\rm rel}$, rather than the planetary 
Einstein crossing time, $\tep$. Eventually, however, when 
$\rho_{\ast,p}\gg 1$ the deviation due to the planet will be completely
suppressed, and the planet will be undetectable.  To quantify these 
trends, we determine the expected $\sn$ of an isolated single-lens event,
\begin{equation}
{{\rm S \over N}} = \frac{N^{1/2}}{\sigma} \left\{\int {\rm d}\tau 
\left[1-A_{fs}(\tau;u_{0,p},\rho_{\ast,p})^{-1}\right]^2\right\}^{1/2}.
\label{eq14}
\end{equation}
where $N$ is the number of measurements per planet Einstein crossing
time $t_{{\rm E},p}$, $\sigma$ is the fractional precision of each
measurement, $A_{fs}$ is the finite-source magnification, and
$\tau\equiv (t-t_0)/t_{{\rm E},p}$ is the time from the peak of the
planetary event in units of the Einstein crossing time.  Although we
assume the magnification is described by single-lens event with no
external shear, our results are approximately applicable to
wide-separations planets as well.  This is because, as we demonstrate
below, planets with mass $m_p\gg M_\oplus$ are detectable with impact
parameters $u_{0,p}\ge 1$, where any deviation arising from the
caustic induced by the primary will be negligible.  On the other hand,
for low-mass planets with $\rho_{\ast,p}\ge 1$, the deviation due to
the caustic vanishes up to fourth order in $\rho_{\ast,p}^{-1}$ when
the source completely encloses the caustic \citep{gould97}.
Therefore, the isolated point-lens approximation should be sufficient
except for a relatively small range near $m_p\sim 0.3~M_\oplus$, 
where we generally underestimate the $\sn$.

The top panel of Figure \ref{fig:six} shows the normalized signal-to-noise 
ratio, $(\sn)/(N^{1/2}/\sigma)$, as a function of the impact parameter 
$u_0$, for source sizes $\rho_{\ast,p}=0.0,2.0,4.0$ and $6.0$.  For 
small sources, $\rho_{\ast,p}\ll 1$, and high-magnification events, 
$u_{0,p}\ll 1$, the term in braces in equation (\ref{eq14}) is 
approximately unity and independent of $u_{0,p}$, and so $\sn \sim N^{1/2}
\sigma^{-1}$.  In the opposite limit of large impact parameter events, 
$u_{0,p}\gg 1$, we find
\begin{equation}
{{\rm S\over N}} = \sqrt{5\pi N\over 4} \sigma^{-1} u_{0,p}^{-7/2}.
\label{eq15}
\end{equation}
The bottom panel of Figure \ref{fig:six} shows the impact parameter 
$(u_{0,p})_{th}$ for which the S/N is equal to the threshold value 
$(\sn)_{th}= (1000)^{1/2}$, for the same assumptions adopted in the 
simulations described in \S\ 3, namely $f_{obs} = 120~{\rm day^{-1}}$, 
and $\sigma=1\%$.  This is the appropriate threshold for detecting 
isolated events in otherwise constant stars \citep{bennett02}.  This 
critical impact parameter $(u_{0,p})_{th}$ corresponds to the cross 
section for the detection of a planet of mass $m_p$.  In the small-source, 
large impact-parameter limit, this can be written
\begin{equation}
(u_{0,p})_{th}\sim 2.3 \left({m_p \over M_\oplus}\right)^{1/14}
\left( {f_{obs}\over 120/{\rm day}} \right)^{1/7} 
\label{eq16}
\end{equation}
$$
\times
\left( {\sigma\over 0.01}\right)^{-2/7}
\left[ {(\sn)_{th} \over\sqrt{10^3}} \right]^{-2/7},
$$
where $f_{obs}$ is the observational frequency.  Therefore, large 
mass-ratio planetary events can generally be detected with planetary 
impact parameters significantly larger than unity.

In the limit of large sources $\rho_{\ast,p}\gg 1$, the magnification 
is strongly affected by finite source effects.  In this case, the 
lightcurve is reasonably well approximated as a boxcar with a duration 
$\sim 2t_\ast$ and an amplitude $A_{fs}\sim 1 + 2 \rho_{\ast,p}^{-2}$ 
\citep{distefano99a,agol03}, implying a $\sn$ of,
\begin{equation}
{{\rm S\over N}}\sim \frac{N^{1/2}}{\sigma} \frac{2^{3/2}
\rho_{\ast,p}^{1/2}}{2+\rho_{\ast,p}^2}.
\label{eq17}
\end{equation}
Therefore, for a fixed $\theta_\ast$ and $\mu_{\rm rel}$, 
planets with a mass less than,
\begin{equation}
m_{p,min}\sim 0.02~M_\oplus 
\left( {f_{obs}\over 120/{\rm d}} \right)
\left( {\sigma\over 0.01}\right)
\left[ {(\sn)_{th} \over\sqrt{10^3}} \right],
\label{eq18}
\end{equation}
cannot be detected due to finite source effects.  This limit is shown 
in the bottom panel of Figure \ref{fig:six}, and corresponds to roughly 
a lunar mass, or a mass ratio of $q \sim 2 \times 10^{-7}$ for a primary 
mass of $\sim 0.3~M_\odot$.  Planets with mass ratio larger than this 
limit, but still well into the finite-source dominated regime, will have 
a detection cross section of $(u_{0,p})_{th}\sim \rho_{\ast,p}$, which 
is generally larger than the corresponding point-source cross section.  
Therefore, for the parameters we have adopted, the cross section as a 
function of decreasing $q$ first decreases until $\rho_{\ast,p}\sim 1$, 
and then increases until $m_{p,min}$, at which point it suddenly plummets.  
See Figure \ref{fig:six}.

\begin{figure}[t]
\epsscale{1.15}
\plotone{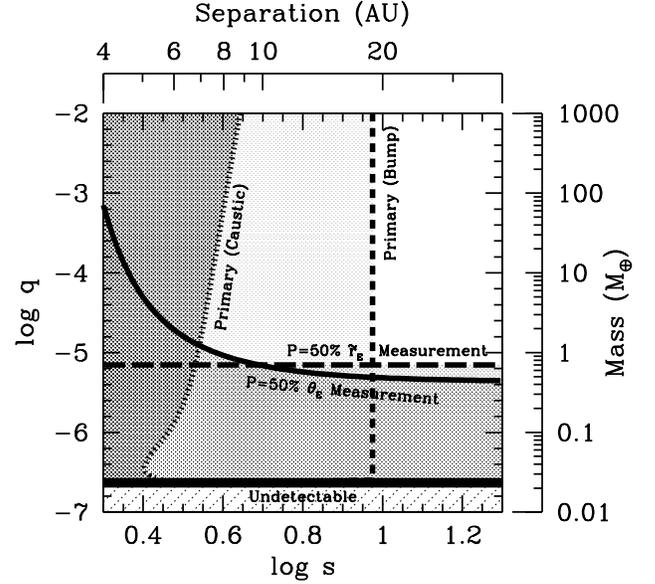}
\caption{\label{fig:seven}
Characterization of wide separation planets as a function of the
planetary separation $s$ and mass ratio $q$ using a next generation
microlensing planet search.  The heavy horizontal line shows the lower
limit on the mass of detectable planets of $m_p\ga 0.02~M_{\oplus}$,
or $q\ga 2\times 10^{-7}$.  In this figure, planets are considered 
detected if they produce deviations with signal-to-noise ratio 
$\sn>(\sn)_{th}=(1000)^{1/2}$.  The solid contour is the upper boundary 
to the region in which $>50\%$ of detectable planetary events will 
yield a $\sim 10\%$ measurement of $\thetae$.  Similarly, the long 
dashed line is the upper boundary to the region for which $>50\%$ of 
detectable planetary events will yield a $\sim 10\%$ measurement of 
$\retilde$.  The dotted contour is the rightmost boundary of the 
region in which $>50\%$ of detectable planets can be identified as 
bound from the deviation caused by the planetary caustic from the 
nominal point-lens form, where $\sn>(160)^{1/2}$ is required for 
detection of the deviation.  Similarly, the dashed vertical line is 
the rightmost boundary of the region in which the magnification of
primary is detectable for $>50\%$ of events with $\sn>(160)^{1/2}$.
The top and right axes show the conversion from $(q,s)$ to
$(m_p,r_\perp)$ assuming a primary mass of $M=0.3~M_\odot$ and
$\re=2~{\rm AU}$.
}\end{figure}

\subsection{Mass Measurement}

We now estimate the fraction of {\it detectable} events that yield
accurate mass measurements, incorporating our estimate for the
cross section for detection from subsection \S\ 4.1.  In
Figure \ref{fig:seven}, we plot the boundary of the region in the
$(m_p,r_\perp)$ or $(q,s)$ plane where $>50\%$ of detectable wide-planet 
events yield a 10\% measurement of $\rho_{\ast,p}$ (thick solid curve).  
Here we have adopted the analytic form for the cross section to measure 
$\rho_{\ast,p}$ to $10\%$ given in equation (\ref{eq13}) with 
${\cal C}_\gamma=2.5$, but normalize to the number of detectable events 
by dividing the resulting probability by $(u_{0,p})_{th}$.  We also show 
in Figure \ref{fig:seven} an approximation to the boundary of the region 
in the $(m_p,r_\perp)$ or $(q,s)$ plane at which we expect $\sim 50\%$ 
of detectable events to yield a 10\% measurement of $\tilde r_{{\rm E},p}$ 
from parallax effects arising from the difference in the lightcurve as 
seen from a satellite at L2 and a ground-based observer (thick long-dashed 
curve).  Here we have assumed that a planet event detected at a 
signal-to-noise ratio $\sn$ yields a fractional error in the projected 
Einstein ring radius of \citep{gould03,han04},
\begin{equation}
\frac{\sigma_{\tilde r_{\rm E}}}{\tilde r_{\rm E}} \sim
\frac{\tilde r_{{\rm E},p}}{d_{sat}} \left({{\rm S\over N}}\right)^{-1},
\label{eq19}
\end{equation}
where $d_{sat}\sim 0.005$ is the projected Earth-satellite separation,
and we have assumed our fiducial value of 
$\pi_{\rm rel}\simeq 42~{\mu{\rm as}}$, so that
$\tilde r_{{\rm E},p} = 0.024~{\rm AU}(m_p/M_\oplus)^{1/2}$.  
Here we have ignored the effect of finite sources on the ability to 
measure $\tilde r_{{\rm E},p}$.  We note that the approximation in 
equation (\ref{eq19}) for the fractional error in $\tilde r_{{\rm E},p}$ 
is very crude; however it captures the primary dependence on $m_p$, and 
agrees reasonably well with more detailed calculations \citep{han04}.

Inspection of Figure \ref{fig:seven} indicates that, by combining
ground-based observations with those of a satellite in an L2 orbit, it
should be possible to measure the mass of a majority of planetary events
arising from planets with $m_p\la M_\oplus$, corresponding to 
$q \la 10^{-5}$, to a fractional precision of $\sim 10\%$.  Mass 
measurements will be possible for a smaller fraction of higher-mass 
planets.  However, since the overall detection rate for such planets will 
likely be higher, a similar or higher number of reasonably-precise mass 
measurements may be possible.

\subsection{Detection of the Primary}

We now consider several methods by which the presence of the primary
star can be detected in isolated events caused by wide planets, 
thus permitting discrimination between bound and isolated planets.

First, as discussed in detail by \citet{han03}, it will be possible 
to identify wide-separation planetary events from the signature of the 
planetary caustic near the peak of light curves.  \citet{han03} found 
that the signatures with $\geq 5\%$ deviation can be detected for 
$\ga 80\%$ of events with $u_{0,p}\le 1$ caused by Jupiter-mass planets 
with  separations $\la 10$ AU, and the probability is still substantial 
for separations even up to $\sim 20$ AU.  Here we consider a simplified 
model for the probability of detecting the primary via the signature of 
the planetary caustic.  We assume that the wide separation can be 
approximated by a Chang-Refsdal lens.  We calculate the fractional 
deviation $\delta$ of the Chang-Refsdal lens from a single lens as a 
function of position along the two axes of symmetry.  We assume that 
a lightcurve with a peak deviation $\delta$ can be distinguished from 
the single-lens case with a signal-to-noise ratio of
\begin{equation}
{{\rm S \over N}} \sim {N^{1/2} \over \sigma} (4\gamma)^{1/2} \delta.
\label{eq20}
\end{equation}
We then determine the semimajor and semiminor axis of the region around
the planetary caustic at which the signal-to-noise ratio is greater than 
a given threshold $(\sn)_{th}$.  The fraction of detectable planetary 
events for which the primary can be detected via the deviation from a 
point lens is then given by equation (\ref{eq12}) normalized by the 
cross-section for detection $u_{0,th}$. We choose $(\sn)_{th}=(160)^{1/2}$.  
This threshold is lower than that assumed for detection of the planetary 
event because there are fewer lightcurves to search for the deviation 
arising from the caustic.  Figure \ref{fig:seven} shows the region in 
the $(q,s)$ plane where $>50\%$ of isolated events arising from bound 
planets give rise to detectable deviations due to the presence of the 
planetary caustic (thick dotted line).  Our results agree well with the 
results of \citet{han03}, when the latter are normalized by 
$(u_{0,p})_{th}$.

Second, wide-separation planetary events can be distinguished by the 
additional long-term bumps in the lightcurve caused by the primary 
star. Compared to the planetary Einstein ring, the Einstein ring of 
the primary star is much larger, implying a larger effective lensing 
region.  Combined with high precision photometry from space observation, 
then, the existence of the primary star can often be noticed even 
without the signatures of the planetary caustic. In the limit of the 
large impact parameter, $u_0$, the ``bump'' due to the primary can be 
detected with signal-to-noise ratio given by equation (\ref{eq15}). Then, 
the cross section to detect the primary-induced bump for a given
threshold signal-to-noise ratio, $(\sn)_{th}$, is given by
\begin{equation}
P=
\cases{
1.0 & for $u_{0,th} > {\hat s}$, \cr
{2\over \pi} \sin^{-1}\left({u_{0,th}\over {\hat s}}\right) & 
for $u_{0,th}<s$,\cr
}
\label{eq21}
\end{equation}
where $u_{0,th}$ is given by equation (\ref{eq16}) with $q=1$.  Figure
\ref{fig:seven} shows the contour (thick short-dashed line) at which 
$P=50\%$ as a function of $(s,q)$, assuming $(\sn)_{th}=(160)^{1/2}$ 
and $M=0.3~{M_\odot}$.   We find that $P=50\%$ for $s \simeq 9.2$, 
corresponding to $r_\perp \simeq 18~{\rm AU}$.  It will be possible to 
detect the primary for essentially all planets with
$s \la 6.52$, or $r_\perp \la 13~{\rm AU}$.

The third method of identifying a wide-separation planet is detecting 
blended light from the host star.  According to \citet{bennett02}, for 
$\sim 1/3$ of events with detected planets from a space lensing mission, 
the planetary host star is either brighter than or within $\sim 2$
magnitudes of the source star's brightness.

Considering the three methods discussed here together, the prospects
for distinguishing isolated planetary events caused by bound and
free-floating planets seem good.  For roughly $1/3$ of all events,
regardless of the mass or separation of the planet, the flux from the
primary should be detectable.  For $>1/2$ of all events caused by wide
planets with $r_\perp \la 5-9~{\rm AU}$ (depending on mass), the
influence of the primary will be detectable in the lightcurve via the
planetary caustic.  Finally, $>1/2$ of all detectable planets with
$r_\perp \ga 20~{\rm AU}$ can be inferred to be bound via the
low-amplitude bump caused by the magnification of the primary.

\section{Conclusion}

With their excellent photometric precision and extremely high temporal
sampling, next-generation microlensing planet searches will be
sensitive to planets with masses almost as low as the moon.  These
searches will employ large-format cameras with large FOV in
order to monitor hundreds of millions of stars simultaneously with
$\sim 10-20$ minute sampling.  Because all stars are monitored
continuously regardless of whether they are being lensed or not, such
searches will be sensitive to isolated events caused by
wide-separation or free-floating planets, in contrast to current
microlensing planet searches.  Such planets are very difficult or
impossible to probe by other planet detection methods.

The scientific return of these wide and free-floating planet
detections would be greatly enhanced if their nature could be
characterized.  In particular, differentiating between wide planets
and free-floating objects is highly desirable, as is the measurement
of their mass.  Generally, the lightcurves of free-floating planets
are grossly similar to those of wide-separation planets.  Furthermore,
microlensing lightcurves generally only yield event time scales, which
are degenerate combinations of the mass, distance, and transverse
velocity of the lens.  However, as recently pointed out by several authors
\citep{gould03,han03,han04}, there are several unique properties of 
next-generation microlensing surveys that should allow better 
characterization of wide-separation and free-floating planets.
Here we have summarized and built on previous works, addressing the
ability of these searches to distinguish wide-separation planets from 
free-floating planets, as well as to measure planet masses.

We have performed detailed simulations of wide-separation planetary
events, and evaluated the probability of measuring the Einstein ring
radius $\thetae$ for these events; this is one of the two additional
quantities needed to measure the lens mass.  From this investigation,
we find that the parameter space of the probability distribution is
divided into two regimes depending on the ratio between the caustic
and normalized source sizes.  In the regime in which the source size
is much larger than the size of the caustic, the probability is not
much different from that of the corresponding free-floating planetary
events.  In the opposite regime in which the caustic size is much
larger than the source size, the probability is significantly higher
than the case without the caustic. As a result, the probability of
$\thetae$ determination for wide-separation planetary events can be
substantially higher than that of free-floating planetary events.  We
find that the effective cross-section of the caustic is about
1.25 times its linear size.

For the majority of events due to planets with mass $m_p\le M_\oplus$, 
it should be possible to measure the angular Einstein ring radius 
$\thetae$ to $\sim 10\%$.  The projected Einstein ring radius $\retilde$ 
should also be measurable to $\sim 10\%$ for the majority of these 
events by combined observations from the ground and a satellite located 
in an L2 orbit.  Thus, it should be possible to measure the mass of 
most wide-separation and free-floating planets of Earth-mass or less.

Finally, we have discussed three methods for distinguishing between
isolated planetary events caused by free-floating and bound planets.
These include detecting the primary through the influence of the
planetary caustic, from the low-amplitude bump in the light curve from
the primary, and from detection of the light from the primary itself.
These three methods should allow one to distinguish between bound and
free-floating planets for $\sim 1/3$ of all events regardless of the
planet separation, $>50\%$ of events with projected separations
$\la 20~{\rm AU}$, and essentially all events with
separations $\la 13~{\rm AU}$.

\acknowledgments 
Work by CH was supported by Astrophysical Research Center for the 
Structure and Evolution of the Cosmos (ARCSEC) of the Korean Science 
\& Engineering Foundation (KOSEF), through the Science Research Center 
(SRC) program, and by JPL contract 1226901.  Work by BSG was supported 
by the Menzel Fellowship from the Harvard College Observatory.  Work 
by JA was supported by a grant from the Leverhume Trust Foundation.
Work by AG was supported by JPL contract 1226901 and NSF grant 02-01266.

\end{document}